\documentclass{article}
\usepackage{spconf}
\usepackage{cite}
\usepackage{adjustbox}
\usepackage[ruled,vlined]{algorithm2e}
\usepackage{amsmath,amssymb,amsfonts}
\usepackage{algorithmic}
\usepackage{graphicx}
\usepackage{textcomp}
\usepackage[colorlinks=false, pdfborder={0 0 0}, breaklinks]{hyperref}
\usepackage{xcolor}
\usepackage{balance}
\usepackage{enumitem}
\usepackage{setspace}
\usepackage{wrapfig}
\usepackage{listings}
\usepackage{color}
\usepackage{caption}
\usepackage{subcaption}
\usepackage{todonotes}

\def\BibTeX{{\rm B\kern-.05em{\sc i\kern-.025em b}\kern-.08em
    T\kern-.1667em\lower.7ex\hbox{E}\kern-.125emX}}

\usepackage{amsmath}
\DeclareMathOperator*{\argmax}{arg\,max}

\usepackage{ctable}

\newcommand{\etal}{\emph{et al.}\xspace}
\newcommand{\ie}{\emph{i.e.}, }
\newcommand{\eg}{\emph{e.g.}, }
\newcommand{\etc}{\emph{etc. }}
\newcommand{\HAS}{\emph{HTTP Adaptive Streaming }}

\newcommand{\VOD}{\emph{Video on Demand }}

\newcommand{\HLS}{\emph{HTTP Live Streaming }}

\newcommand{\QoE}{\emph{Quality of Experience}\xspace}

\newcommand{\jaslad}{\texttt{JASLA}\xspace}

\begin{document}

\title{Just Noticeable Difference-aware\\Per-Scene Bitrate-laddering for Adaptive Video Streaming}
\name{
\begin{tabular}{@{}c@{}}
Vignesh V Menon$^1$ \qquad 
Jingwen Zhu$^2$ \qquad 
Prajit T Rajendran$^3$ \qquad  
Hadi Amirpour$^1$ \\
Patrick Le Callet$^{2}$ \qquad 
Christian Timmerer$^{1}$
\end{tabular}\vspace{-0.5em}}

\address{\small $^1$Christian Doppler Laboratory ATHENA, Alpen-Adria-Universit{\"a}t, Klagenfurt, Austria \\
\small $^2$Nantes Universite, Ecole Centrale Nantes, CAPACITES SAS, CNRS, LS2N, UMR 6004, F-44000 Nantes, France\\
\small $^3$CEA, List, F-91120 Palaiseau, Université Paris-Saclay, France
}

\maketitle

\begin{abstract}
In video streaming applications, a fixed set of bitrate-resolution pairs (known as a \textit{bitrate ladder}) is typically used during the entire streaming session. However, an optimized bitrate ladder per scene may result in \textit{(i)} decreased storage or delivery costs or/and \textit{(ii)} increased \QoE. This paper introduces a Just Noticeable Difference (JND)-aware per-scene bitrate ladder prediction scheme (\jaslad) for adaptive video-on-demand streaming applications. \jaslad predicts jointly optimized resolutions and corresponding constant rate factors (CRFs) using spatial and temporal complexity features for a given set of target bitrates for every scene, which yields an efficient constrained Variable Bitrate encoding. Moreover, bitrate-resolution pairs that yield distortion lower than one JND are eliminated. 
Experimental results show that, on average, \jaslad yields bitrate savings of 34.42\% and 42.67\% to maintain the same PSNR and VMAF, respectively, compared to the reference \HLS (HLS) bitrate ladder Constant Bitrate encoding using x265 HEVC encoder, where the maximum resolution of streaming is Full HD (1080p). Moreover, a 54.34\% average cumulative decrease in storage space is observed.
\end{abstract}

\begin{keywords}
Bitrate ladder, per-scene encoding, video streaming, Just Noticeable Difference.
\end{keywords}
\vspace{-0.5em}
\section{Introduction}
\vspace{-0.75em}
\textbf{\textit{Motivation:}} \VOD (VoD) and live video streaming are widely embraced in video services, and their applications have attracted tremendous attention in recent years~\cite{index2019cisco}. 
Since streaming services continuously adapt video delivery to the end user's network conditions and device capabilities, \HAS (HAS) continues to grow and has become the \textit{de-facto} standard for delivering video over the Internet~\cite{Bentaleb2019}. In HAS, each video is encoded at a set of bitrate-resolution pairs, referred to as \emph{bitrate ladder}. Traditionally, a fixed bitrate ladder, \eg \HLS (HLS) bitrate ladder\footnote{\label{apple_hls_ref}\href{https://developer.apple.com/documentation/http\_live\_streaming/hls\_authoring\_specification\_for\_apple\_devices}{https://developer.apple.com/documentation/http\_live\_streaming/ hls\_authoring\_specification\_for\_apple\_devices}, last access: Apr 30, 2023.}, is used for all video contents. However, due to the vast diversity in video content characteristics and network conditions, the ``one-size-fits-all'' can be optimized per \textit{scene} to increase the \QoE (QoE) or decrease the bitrate of the representations as introduced for VoD services~\cite{netflix_paper}. 
\begin{figure}[t]
    \centering
    \includegraphics[clip,width=0.35\textwidth]{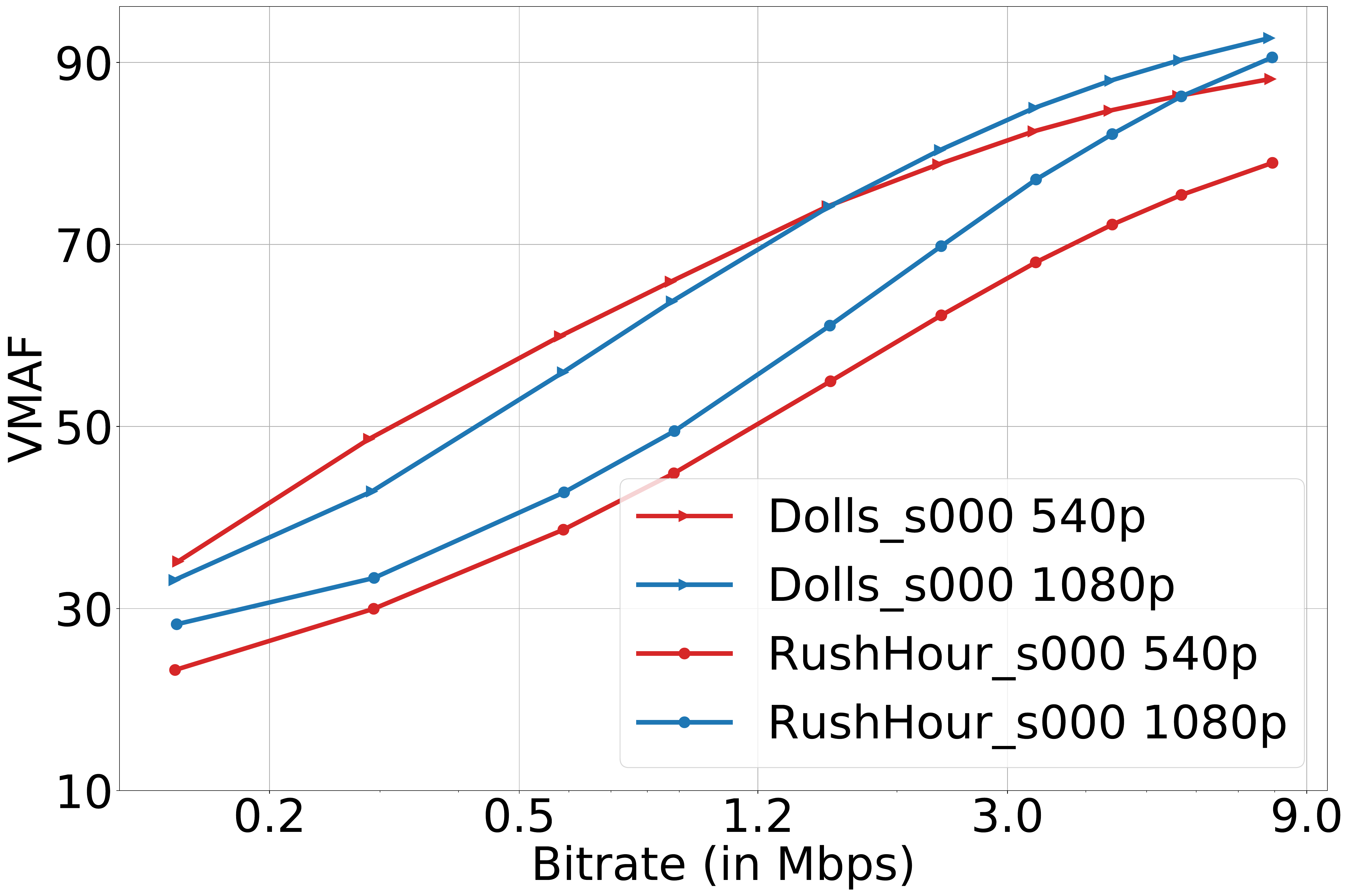}
\vspace{-0.3em}
\caption{RD curve of 540p and 1080p CBR encodings of \textit{Dolls\_s000} and \textit{RushHour\_s000}~\cite{VCD_ref} video sequences using x265 HEVC encoder at \textit{slower} preset.}
\vspace{-1.0em}
\label{fig:convex_hull_motive}
\end{figure}
Per-scene encoding schemes are based on the fact that one resolution performs better than others in a \textit{scene} for a given bitrate range, and these regions depend on the \textit{video complexity}~\cite{jtps_ref}. As shown in Fig.~\ref{fig:convex_hull_motive}, for \textit{Dolls\_s000}, the cross-over bitrate between 540p and 1080p resolutions happens at approximately 2.0 Mbps, which means at bitrates lower than 2.0 Mbps, 540p resolution outperforms 1080p in terms of VMAF\footnote{\label{ref_vmaf}\href{https://netflixtechblog.com/vmaf-the-journey-continues-44b51ee9ed12}{https://netflixtechblog.com/vmaf-the-journey-continues-44b51ee9ed12}, last access: Apr 30, 2023.}. In comparison, at bitrates higher than 2.0 Mbps, 1080p resolution outperforms 540p. On the other hand, for \textit{RushHour\_s000}, 1080p yields higher VMAF over the entire bitrate range, which means 1080p should be selected for the bitrate ladder for the entire bitrate range. This \textit{content-dependency} to choose the optimal bitrate-resolution pairs is the basis for introducing a \textit{per-scene} encoding scheme. 
Each scene in a video and its corresponding downscaled versions are encoded at several bitrates. The bitrate-resolution pair with the highest quality is selected for each target bitrate~\cite{netflix_paper}. Considering $M$ resolutions and $N$ bitrates, $M\times N$ test encodings are needed to determine the optimal per-scene bitrate ladder. To avoid a brute force encoding of all bitrate-resolution pairs, some methods pre-analyze the video contents\footnote{\href{https://bitmovin.com/per-title-encoding/}{https://bitmovin.com/per-title-encoding/}, last access: Apr 30, 2023.}. Katsenou~\etal~\cite{gnostic} introduced a content-gnostic method that employs machine learning to find the bitrate range for each resolution that outperforms other resolutions. Bhat~\etal~\cite{res_pred_ref1} proposed a Random Forest (RF) classifier to decide the best encoding resolution over different quality ranges.

\begin{figure}[t]
\centering
\includegraphics[clip,width=0.35\textwidth]{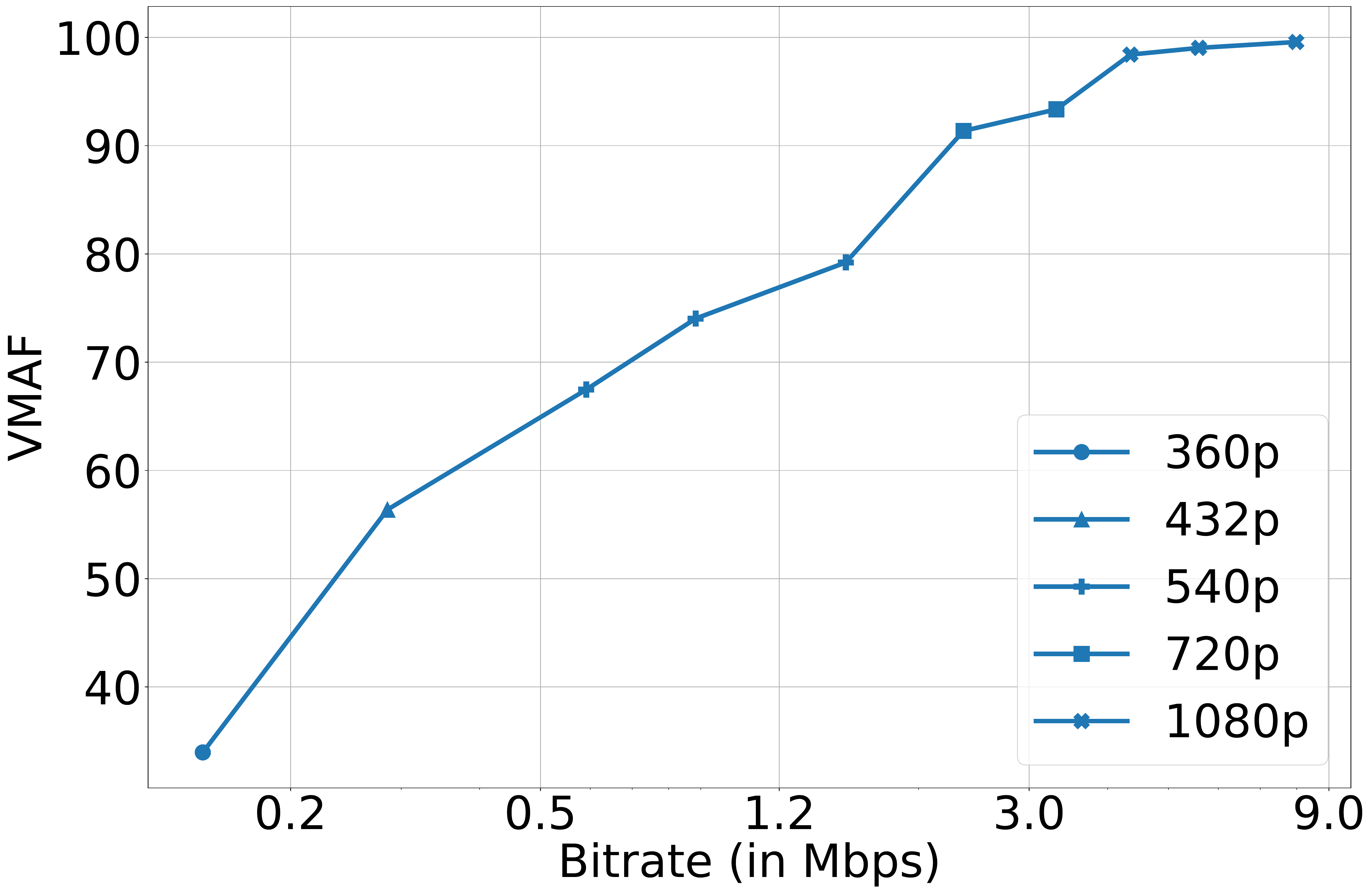}
\vspace{-0.4em}
\caption{RD curve of HLS$^{\ref{apple_hls_ref}}$ CBR encoding of \textit{Characters\_s000} video sequence (segment) of VCD dataset~\cite{VCD_ref}  using x265 HEVC encoder at \textit{slower} preset. The points with a bitrate greater than 3.6 Mbps are in the perceptually lossless region.}
\label{fig:motive}
\vspace{-0.86em}
\end{figure}

\begin{figure*}[t]
\begin{center}
\includegraphics[trim={0cm 5.6cm 0.0cm 0cm}, clip, width=0.9995\linewidth]{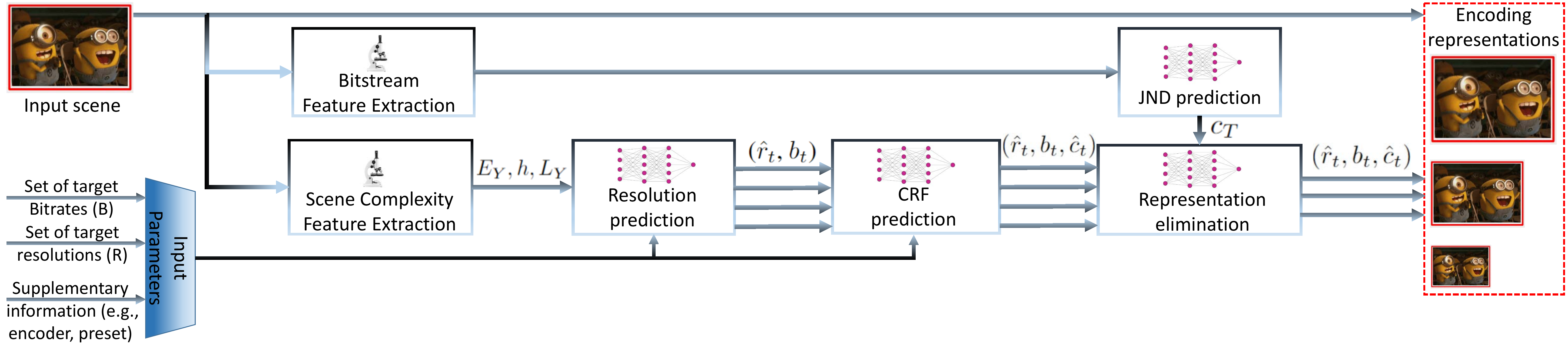}
\end{center}
\vspace{-1.75em}
\caption{\jaslad architecture.}
\vspace{-1.25em}
\label{fig:jasla_enc}
\end{figure*}

As shown in Fig.~\ref{fig:motive}, the bitrate-resolution pairs of the bitrate ladder may not always be perceptually different in video quality. It is observed that there are multiple representations with similar video quality (\ie VMAF close to 100) for the \textit{Characters\_s000} sequence using the HLS bitrate ladder. Having many perceptually redundant representations for the bitrate ladder may not result in improved quality of experience, but it may lead to increased storage and bandwidth costs~\cite{rep_num_ref}. Hence, predicting the threshold where visible distortion occurs compared to the original scene (referred to as JND) is critical. Wang~\etal~\cite{jnd_ref1} proposed a model using Support Vector Regression (SVR) to predict JND based on masking effect features~\cite{mask_effect_ref} extracted from the source video and quality degradation features computed from various encoded videos. Zhang~\etal~\cite{jnd_pred2} improved the JND prediction accuracy by considering the spatial and temporal information features via deep learning. However, the source (raw) video needs to be encoded several times (\eg QP from 1 to 51 with a step size of 1) before conducting these models, which is computationally expensive. Zhu~\etal~\cite{jnd_ref2} proposed a JND prediction model that only inputs the source video. Moreover, the Video-Wise JND dataset (HD-VJND) is collected using CRF as a proxy, and a JND prediction model is proposed by extracting three types of features~\cite{zhu2022subjective}. Even though these JND prediction models only take the source video as input, they are still computationally intensive because of the high complexity of features (\eg masking effect features~\cite{mask_effect_ref}).

\textbf{\textit{Contributions:}} In this paper, a \textbf{J}ust Noticeable Difference (JND)-\textbf{a}ware per-\textbf{s}cene bitrate \textbf{la}dder prediction scheme (\jaslad) is proposed that improves encoding bitrate ladders for adaptive video-on-demand streaming applications. \jaslad predicts optimized resolution and the corresponding constant rate factor (CRF) using spatial and temporal complexity features, for all target bitrates defined by the streaming service provider for efficient constrained Variable Bitrate (cVBR) encoding. Furthermore, a JND threshold prediction scheme is implemented to eliminate the representations which yield distortion lower than the noticeable distortion for every scene. 
\vspace{-0.7em}
\section{\jaslad Architecture}
\label{sec:ppte_framework}
\vspace{-0.6em}
The \jaslad architecture is shown in Fig.~\ref{fig:jasla_enc}. The resolution and the corresponding CRF for each bitrate in the bitrate ladder are predicted for every scene using the scene's spatial and temporal complexity features, the set of pre-defined resolutions ($R$), and the set of pre-defined bitrates ($B$) for an efficient cVBR steaming. An optimized bitrate ladder for every scene ensures streaming quality with no bitrate fluctuations. $R$ is input to \jaslad to confirm that only the resolutions supported by the streaming service provider are selected to generate the optimized bitrate ladder. Next, the bitrate-resolution pairs whose perceptual quality is less than one JND compared to the source video are eliminated. In this way, the number of representations needed for streaming is reduced. The encoding process is carried out only for the predicted bitrate-resolution-CRF pairs for every scene. 

\jaslad comprises three steps: \textit{(i)} scene complexity features extraction, \textit{(ii)} optimized resolution and CRF prediction, and \textit{(iii)} JND threshold prediction which are described in the following.
\vspace{-0.6em}
\subsection{Scene Complexity Features Extraction}
\label{sec:feature_extraction}
\vspace{-0.3em}
In video streaming applications, an intuitive method for feature extraction would be to utilize Convolutional Neural Networks (CNNs)~\cite{3d_cnn_vqa_ref}. However, such models have several inherent disadvantages, such as higher training time, inference time, and storage requirements, which are impractical in streaming applications. 
The popular state-of-the-art video complexity features are Spatial Information (SI) and Temporal Information (TI)\footnote{\label{ref_siti}\href{https://www.itu.int/rec/T-REC-P.910-202207-I}{https://www.itu.int/rec/T-REC-P.910-202207-I}, last access: Apr 30, 2023.}. But the correlation of SI and TI features with the encoding output features such as bitrate, encoding time \etc are very low, which is insufficient for encoding parameter prediction in streaming applications~\cite{mmsp_paper_ref, ppte_ref, vqa_mhv_ref, vca_ref}.

In this paper, seven DCT-energy-based features~\cite{dct_ref}, the average luma texture energy $E_{Y}$, the average gradient of the luma texture energy $h$, the average luminescence $L_{Y}$, the average chroma texture energy $E_{U}$ and $E_{V}$ (for U and V planes) and the average chrominance $L_{U}$ and $L_{V}$ (for U and V planes), which are extracted using VCA\footnote{\label{ref_vca}\href{https://vca.itec.aau.at}{https://vca.itec.aau.at}, last access: Apr 30, 2023.} open-source video complexity analyzer~\cite{ds_paper_ref,vca_ref} are used as the spatial and temporal complexity measures~\cite{jtps_ref,vqa_icip_ref} of every scene.  
\subsection{Optimized Resolution and CRF Prediction}
\label{sec:res_pred}
\vspace{-1.52em}
For each scene, the optimized resolution for a given target bitrate is predicted using the scene's spatial and temporal features, the set of supported resolutions ($R$), and the set of target bitrates ($B$). To determine the bitrate-resolution pairs of the bitrate ladder, VMAF is predicted for each target bitrate ($b_t$) in the set $B$ for all resolutions $\Tilde{r}$ in $R$, denoted as $v_{\Tilde{r}, b_{t}}$. From the predicted VMAF values, the resolution which yields the maximum VMAF value is chosen as the optimized resolution for the target bitrate. Random Forest (RF) models are trained to predict VMAF for every resolution supported by the streaming service provider. This ensures \textit{scalability} of design, where there is no requirement to retrain the entire network to add a new resolution to the framework.

\begin{algorithm}[t]
\textbf{Inputs:}\\
\quad $R$~: set of all resolutions $\Tilde{r}_m~\forall~m \in [1, M]$ \\
\quad $M$~: number of resolutions in $R$\\
\quad $B$~: set of all bitrates $b_t~\forall~t \in [1, N]$ \\
\quad $N$~: number of bitrates in $B$\\
\quad $E_Y, h, L_Y$~: average scene complexity \\
\textbf{Output:} $(\hat{r},b,\hat{c})$ pairs of the bitrate ladder \\

\For{$t \in [1, N]$}
{
$\quad$ \For{$m \in [1, M]$}{
$\quad$ Determine $v_{\Tilde{r}_{m}, b_{t}}$ with $[E_Y, h, L_Y, log(b_t)]$, using the model trained for $\Tilde{r}_{m}$. \\
}
$\quad$ $\hat{r}_{t} = \argmax_{\Tilde{r}_{m}\in R}(v_{\Tilde{r}, b_{t}})$\\
$\quad$ Determine $\hat{c}_{t}$ with $[E_Y, h, L_Y, log(b_t)]$, using the model trained for $\hat{r}_{t}$.\\
$\quad (\hat{r}_{t}, b_{t}, \hat{c}_{t})$ is the $(t)^{th}$ point of the bitrate ladder\\
}
\caption{Optimized resolution and CRF prediction}
\label{algo:res_pred}
\end{algorithm}
\setlength{\textfloatsep}{3pt}
\begin{figure*}[t]
\centering
\includegraphics[trim={3cm 3cm 0.2cm 1cm}, clip, width=0.880\linewidth]{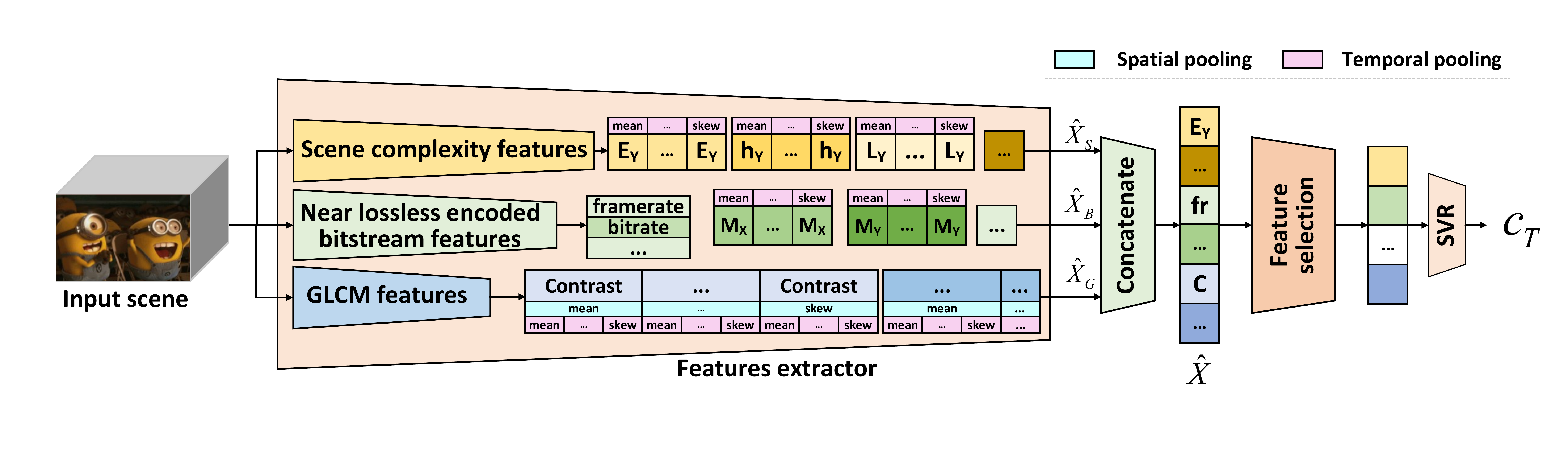}
\vspace{-0.2cm}
\caption{JND threshold prediction model architecture.}
\label{fig:jnd_model}
\vspace{-0.95em}
\end{figure*}

Using the $E_Y$, $h$, $L_Y$ features, optimized CRF $\hat{c_{t}}$ is estimated for every ($\hat{r}_{t}$, $b_{t}$) representation of the bitrate ladder for cVBR encoding. Prediction models are trained for each resolution $\Tilde{r}$ in $R$, which determines $\hat{c_{t}}$ based on $E_Y$, $h$, $L_Y$ and $log(b_{t})$ for every scene. The minimum and maximum CRF ($c_{min}$ and $c_{max}$, respectively) are chosen based on the target codec. For example, x265$^{\ref{ref_x265}}$ supports a CRF range between 0 and 51. The prediction algorithm for the bitrate ladder is shown in Algorithm~\ref{algo:res_pred}.
\vspace{-0.70em}
\subsection{JND Threshold Prediction}
\label{sec:vmaf_thresh}
\vspace{-0.22em}
This paper proposes a reduced complexity JND prediction model derived from \cite{zhu2022subjective}, which predicts the minimum CRF where perceptual distortion is introduced, as shown in Fig.~\ref{fig:jnd_model}. Three types of low-complexity features, \textit{(i)} scene complexity features, \textit{(ii)} bitstream features, and \textit{(iii)} Gray-Level Co-occurrence Matrix (GLCM) features are extracted from the input video scene to predict the JND threshold CRF ($c_T$). 
 \begin{enumerate}[label=\arabic*),leftmargin=*,nosep]
\item \textit{Scene complexity features} ${X_S}$ = \{ ${E_Y}$, $h$, ${L_Y}$, ${E_U}$, ${E_V}$, ${L_U}$, ${L_V}$\} are used instead of the masking effect features in \cite{zhu2022subjective} due to efficiency reasons. 
\item \textit{Bitstream features}\cite{bitstream_analyze_ref}: The scene is first compressed into a near-lossless version to extract bitstream features (denoted as $X_B$), including framerate, bitrate, framesize, motion (horizontal and vertical), \etc The bitstream features are extracted using \textit{videoparse}\footnote{\href{https://github.com/Telecommunication-Telemedia-Assessment/bitstream_mode3_videoparser}{https://github.com/Telecommunication-Telemedia-Assessment/bitstream\_mode3\_videoparser}, last access: Apr 30, 2023.} without decoding pixel information~\cite{bitstream_analyze_ref1}.
\item \textit{Gray-Level Co-occurrence Matrix (GLCM) features}~\cite{GLCM_ref}: Among the nature scene statistic features used in ~\cite{zhu2022subjective}, only GLCM features~\cite{GLCM_ref} are used in this paper owing to their importance in JND prediction. Each frame is cropped into patches of size $Q\times Q$. For each patch, the co-occurrence matrix is computed as ${X_{G}}$ = \{contrast, dissimilarity, homogeneity, angular second moment, energy, correlation\}.
 \end{enumerate}
\vspace{-0.30em}
This paper considers five types of pooling, \ie \{ mean, std, max, skew, kurt\}. Pooled scene complexity features is computed as ${{\hat X}_S}={F_t}({X_S})$, where $F_t$ is the temporal pooling among frames. Similarly, pooled bitstream features are estimated as ${{\hat X}_G}={F_t}({X_G})$. Pooled GLCM features are computed as ${{\hat X}_G}= {F_t}({F_s}({X_G}))$, where $F_s$ is the spatial pooling among patches.

All extracted features are concatenated into one feature vector, and Forward-Sequential Feature Selection (F-SFS)~\cite{ferri1994comparative} selects 15 features. The number of features is determined based on a trade-off between complexity and accuracy. The selected features are shown in Table~\ref{tab:15_feature}. These features are fed into a Support Vector Regression (SVR) for predicting the minimum CRF ($c_T$) where noticeable quality distortion (first JND) is observed.

\textit{Representation elimination}: $c_T$ is used to eliminate perceptually redundant representations from the bitrate ladder as shown in Algorithm~\ref{algo:res_eliminate}. There shall be only one representation in the bitrate ladder where the selected optimized resolution is the maximum supported resolution ($r_{max}$), and the predicted optimized CRF is lower than $c_T$. Other higher bitrate representations are eliminated.

\begin{table}[t]
\caption{List of the fifteen features selected by F-SFS.}
\vspace{-0.4em}
\centering
\resizebox{0.485\textwidth}{!}{
\begin{tabular}{l |c |c }
\specialrule{.12em}{.05em}{.05em}
${{\hat X}_S}={F_t}({X_S})$ & $\hat{X}_B={F_t}({X_B})$ & $\hat{X}_G$ = ${F_t}({F_s}({X_G}))$ \\
\specialrule{.12em}{.05em}{.05em}
max($L_Y$) & kurt(AvMotionX)         & mean(mean(dissimilarity)) \\
max($L_U$) & kurt(AvMotionY)         & kurt(kurt(dissimilarity)) \\
           & kurt(SpatialComplexity) & max(mean(homogeneity)) \\
           &                         & mean(mean(homogeneity)) \\
           &                         & skew(std(angular second moment)) \\
           &                         & kurt(std(angular second moment)) \\
           &                         & kurt(skew(angular second moment)) \\
           &                         & mean(skew((energy)) \\
           &                         & std(max((correlation)) \\           
           &                         & kurt(max((contrast)) \\            
\specialrule{.12em}{.05em}{.05em}
\end{tabular}}
\label{tab:15_feature}
\end{table}
\vspace{-0.85em}

\section{Evaluation}
\label{sec:evaluation}
\vspace{-0.65em}
\subsection{Test Methodology}
\label{sec:test_methodology}
\vspace{-0.3em}
In this paper, four hundred video sequences (\ie 80\% of all sequences) from the Video Complexity Dataset~\cite{VCD_ref} are used as the training dataset, and the remaining (20\%) is used as the test dataset. The video sequences are encoded at 30fps using x265\footnote{\label{ref_x265}\href{https://videolan.org/developers/x265.html}{https://videolan.org/developers/x265.html}, last access: Apr 30, 2023.} v3.5 with the \textit{slower} preset. The bitrate-ladder specified in Apple HLS authoring specifications$^{\ref{apple_hls_ref}}$ are considered in the evaluation, \ie R= \{360p, 432p, 540p, 720p, 1080p\} and $B$ = \{145, 300, 600, 900, 1600, 2400, 3400, 4500, 5800, 8100\}. $E_Y$, $h$ and $L_Y$ features are extracted using VCA$^{\ref{ref_vca}}$ v1.5 open-source video complexity analyzer~\cite{vca_ref} run in eight CPU threads using x86 SIMD optimization~\cite{x86_simd_ref}. Hyperparameter tuning is performed to obtain a balance between the model size and performance for VMAF and CRF prediction models, which results in the following parameters\footnote{\href{https://scikit-learn.org/stable/modules/ensemble.html\#forests-of-randomized-trees}{https://scikit-learn.org/stable/modules/ensemble.html\#forests-of-randomized-trees}, Last access: Apr 30, 2023.} for VMAF and CRF prediction models: \textit{min\_samples\_leaf} = 1, \textit{min\_samples\_split} = 2, \textit{n\_estimators} = 100, and \textit{max\_depth} = 14.
Furthermore, the bitstream features are extracted from the CRF=5 encoded bitstream for each scene. $Q\times Q$ is set as $64\times64$ to determine GLCM features. The JND prediction model is trained on HD-VJND datasets~\cite{zhu2022subjective} for the Full HD (1080p) resolution by five-fold cross-validation. The kernel of SVR is the Radial basis function\footnote{\href{https://scikit-learn.org/stable/modules/generated/sklearn.svm.SVR.html}{https://scikit-learn.org/stable/modules/generated/sklearn.svm.SVR.html}, Last access: Apr 30, 2023.} with the parameters $\epsilon=0.0001$ and regularization parameter $C=0.1$ determined by a greedy hyperparameter search.

\begin{algorithm}[t]
\textbf{Inputs:}\\
\quad $N$~: number of bitrates in $B$\\
\quad $(\hat{r},b, \hat{c})$ pairs of the bitrate ladder \\
\quad $c_T$~: JND threshold CRF \\
\quad $r_{max}$~: maximum resolution in $R$ \\
\textbf{Output:} $(\hat{r},b,\hat{c})$ pairs for encoding \\
$t=1, flag=0$\\
\While{$t \leq N$}
{
\If{$\hat{r}_{t} == r_{max}$ \text{and} $\hat{c}_{t} < c_{T}$}
{$flag++$}
\If{flag $>$ 1}
{Eliminate $(\hat{r}_{t},b_{t}, \hat{c}_{t})$ from the ladder.}
$t++$
}
\caption{Representation elimination}
\label{algo:res_eliminate}
\end{algorithm}
\setlength{\textfloatsep}{4pt}

\begin{figure*}[!t]
\centering
\begin{subfigure}{0.229\textwidth}
    \centering
   \includegraphics[clip,width=\textwidth]{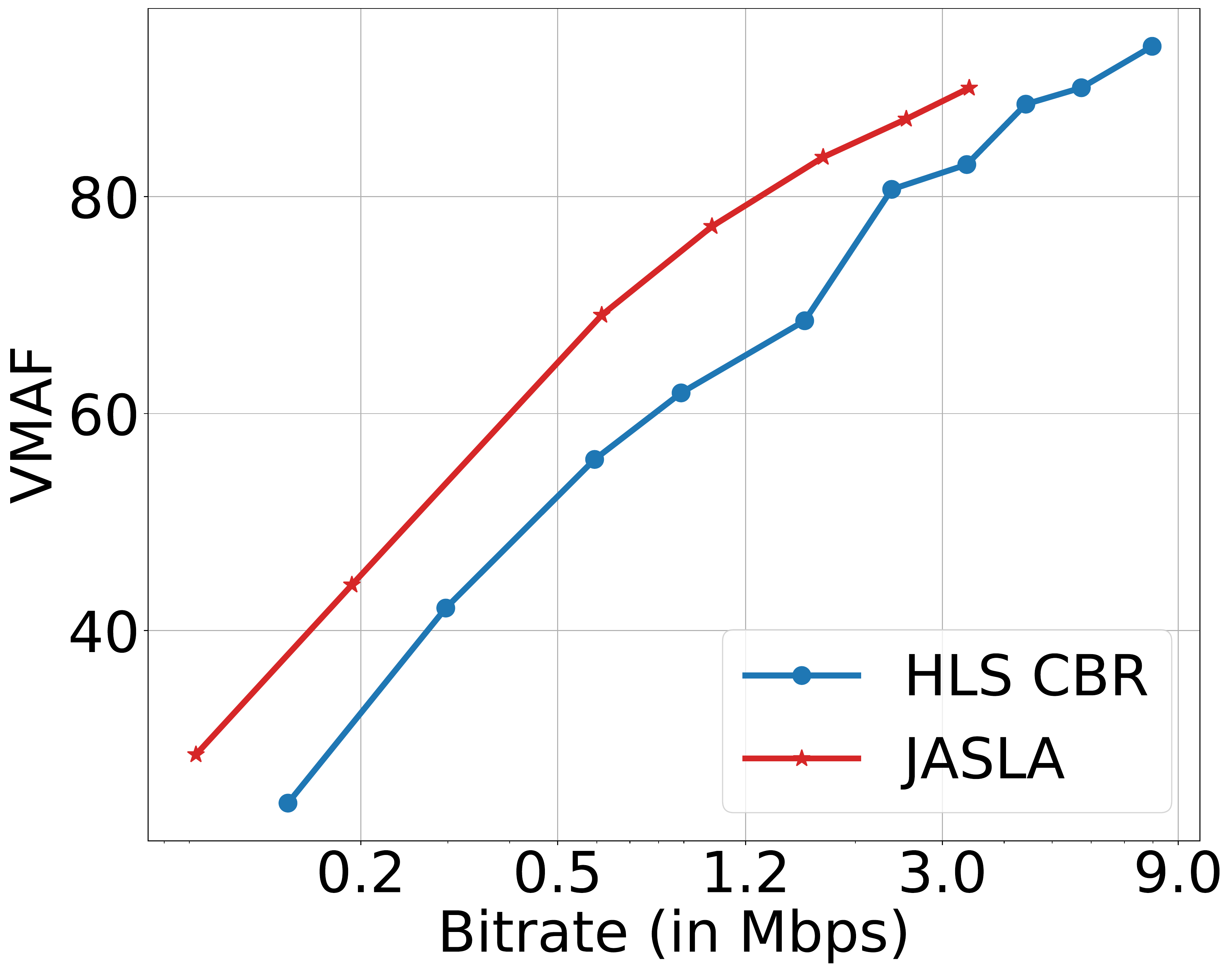}
    \caption{\textit{Bunny\_s000}}
\end{subfigure}
\hfill
\begin{subfigure}{0.229\textwidth}
    \centering
   \includegraphics[clip,width=\textwidth]{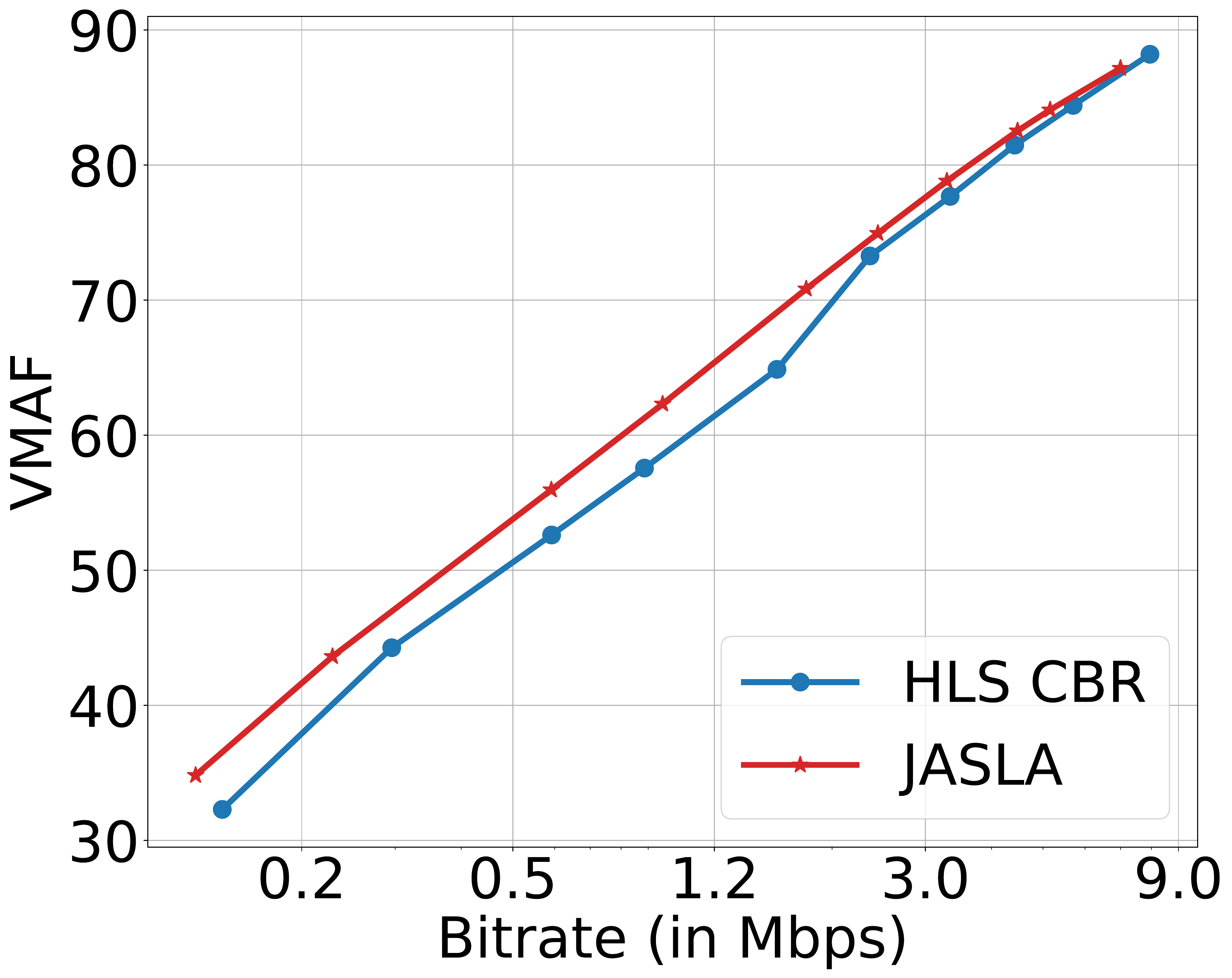}
    \caption{\textit{Bosphorus\_s000}}
    \label{fig:bosphorus_rd1}
\end{subfigure}
\hfill
\begin{subfigure}{0.229\textwidth}
    \centering
   \includegraphics[clip,width=\textwidth]{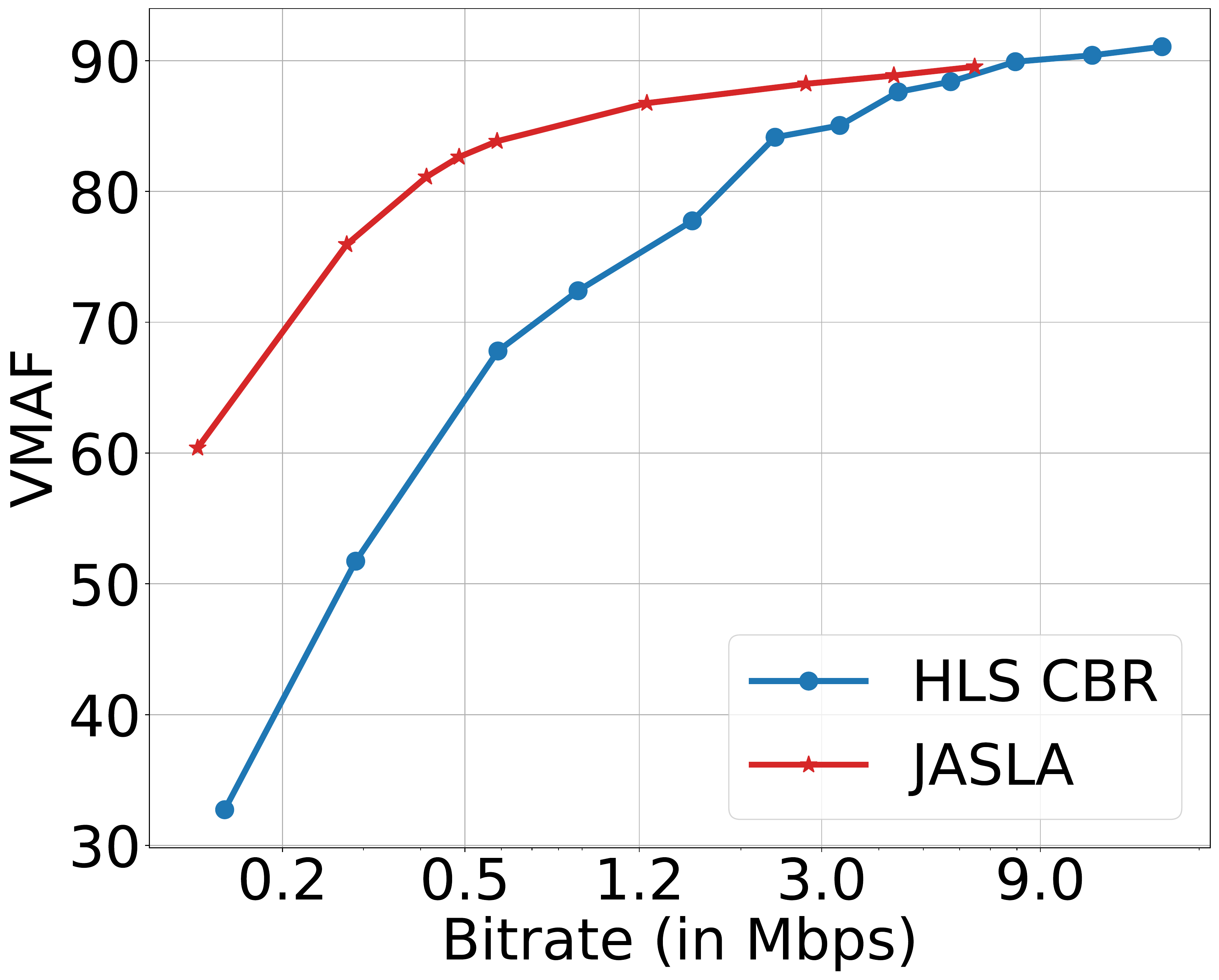}
    \caption{\textit{HoneyBee\_s000}}
    \label{fig:honeybee_rd1}
\end{subfigure}
\hfill
\begin{subfigure}{0.229\textwidth}
    \centering
   \includegraphics[clip,width=\textwidth]{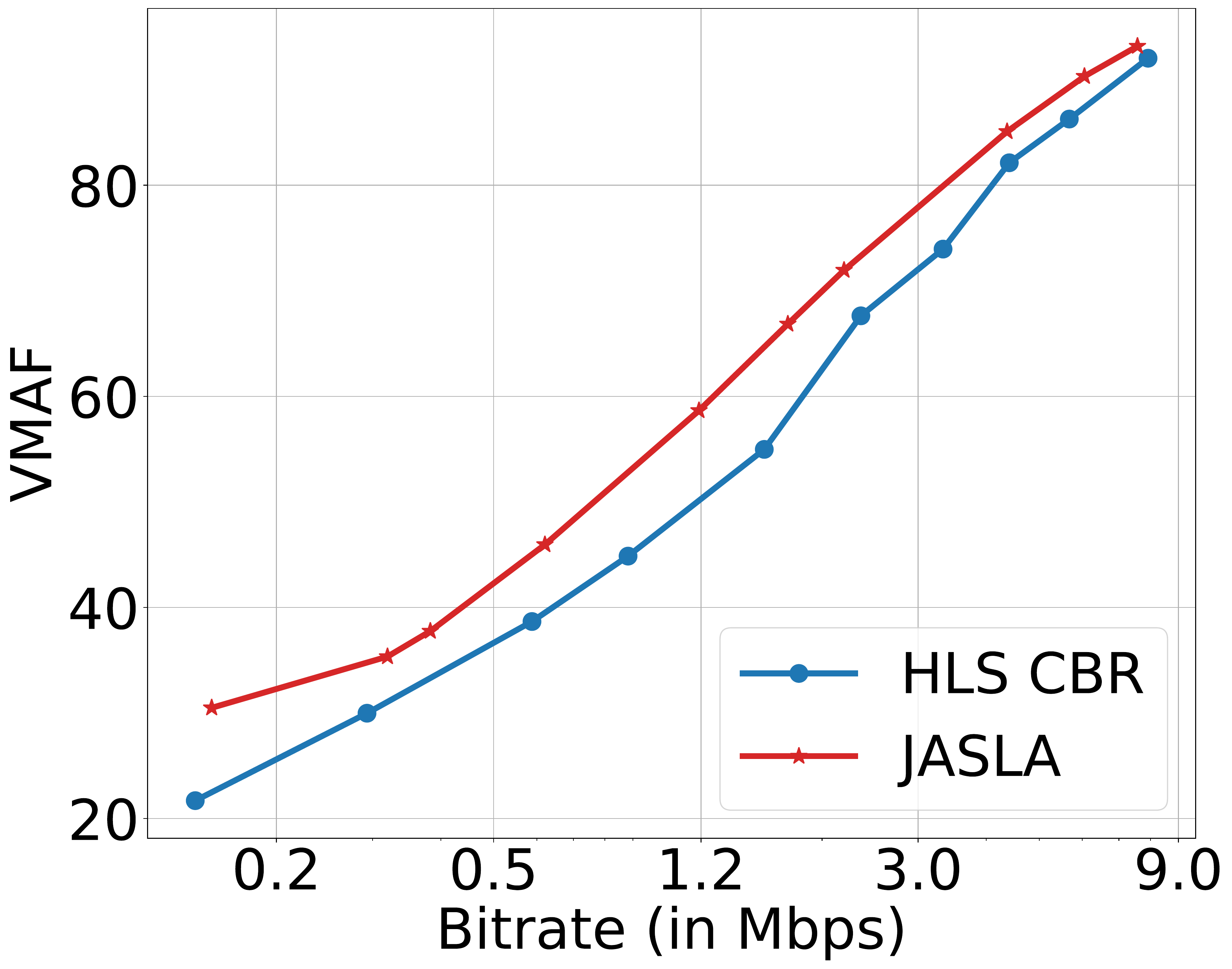}
    \caption{\textit{RushHour\_s000}}
    \label{fig:rushhour_rd1}
\end{subfigure}
\vfill
\begin{subfigure}{0.229\textwidth}
    \centering
   \includegraphics[clip,width=\textwidth]{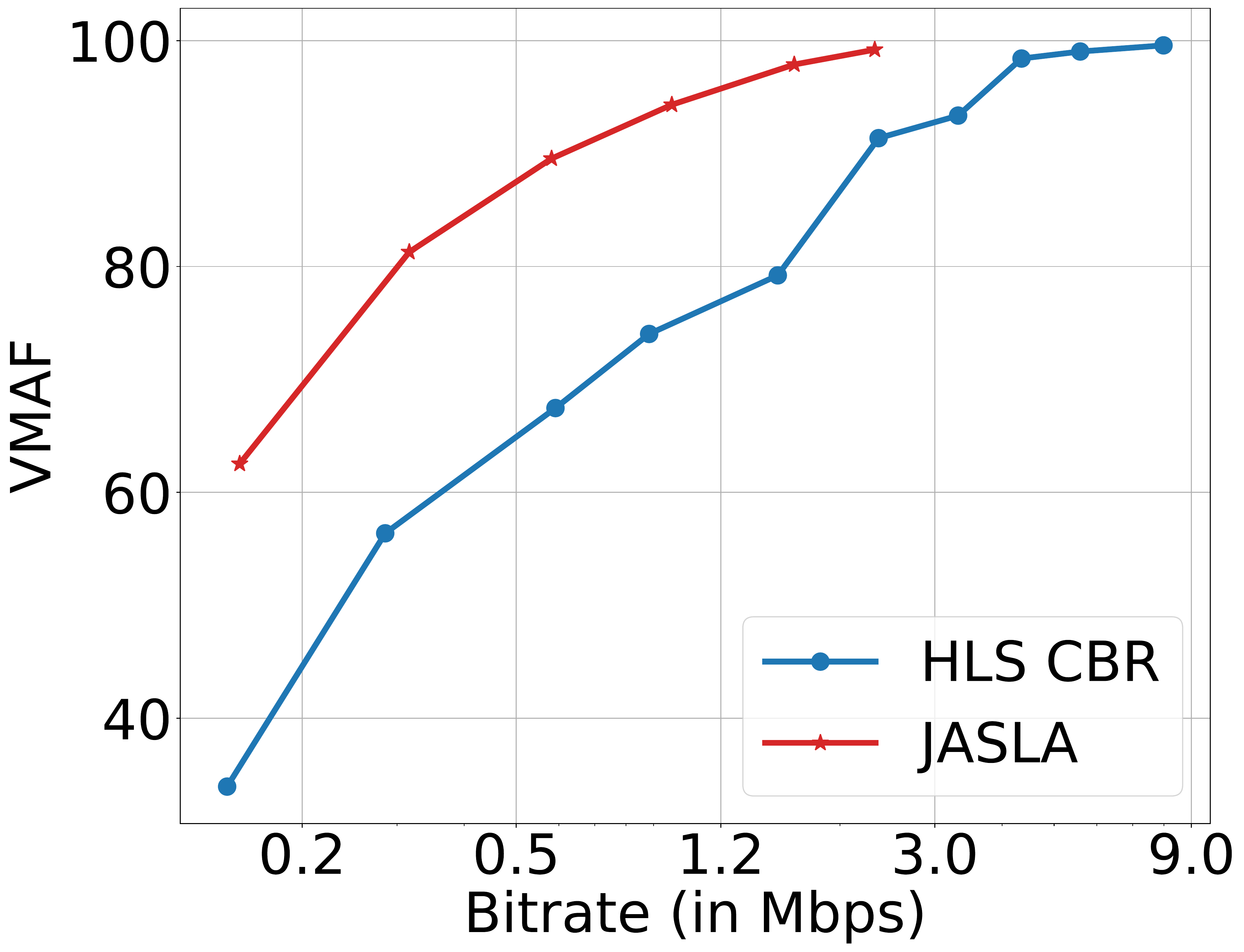}
    \caption{\textit{Characters\_s000}}
\end{subfigure}
\hfill
\begin{subfigure}{0.229\textwidth}
    \centering
   \includegraphics[clip,width=\textwidth]{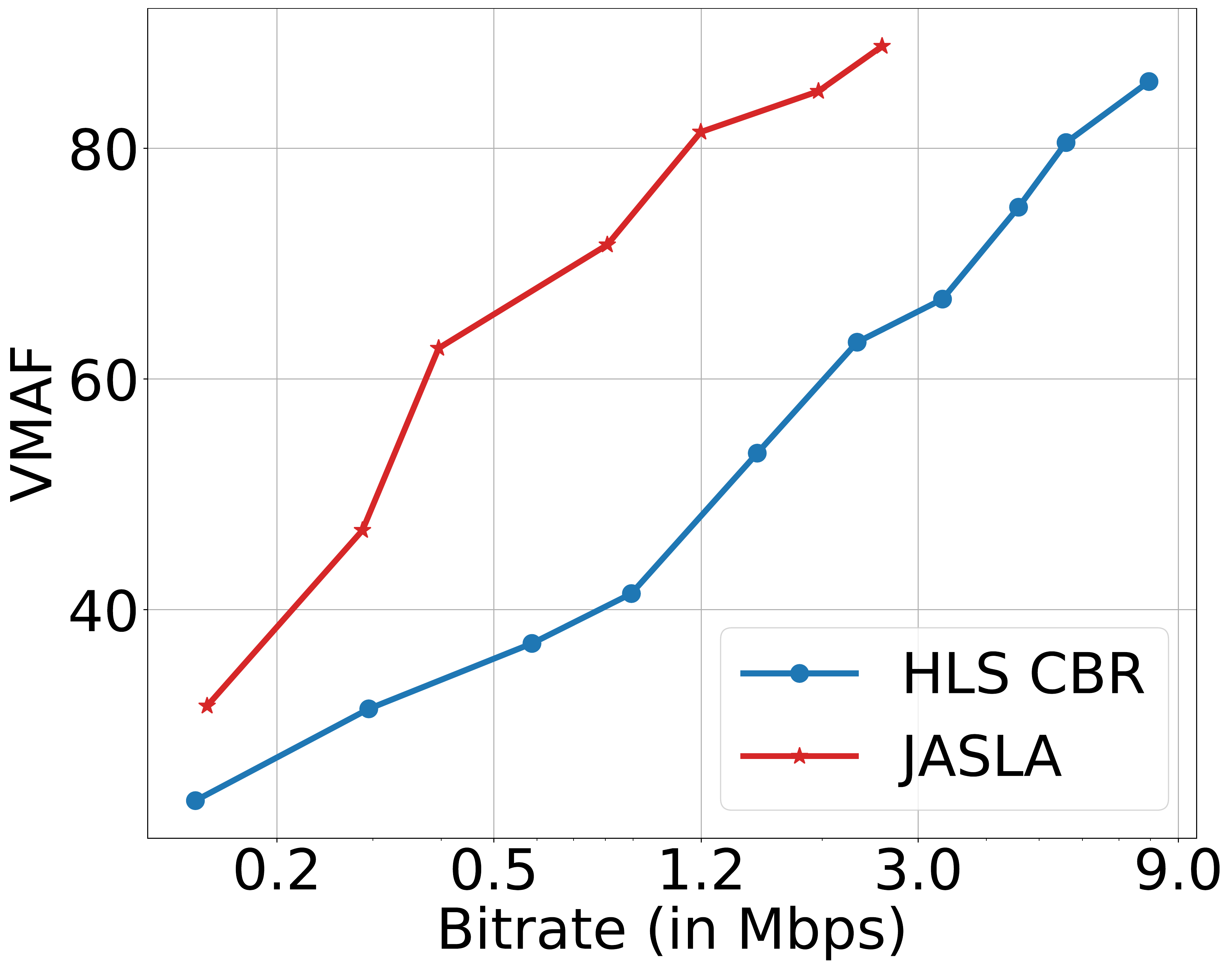}
    \caption{\textit{Eldorado\_s005}}
\end{subfigure}
\hfill
\begin{subfigure}{0.229\textwidth}
    \centering
   \includegraphics[clip,width=\textwidth]{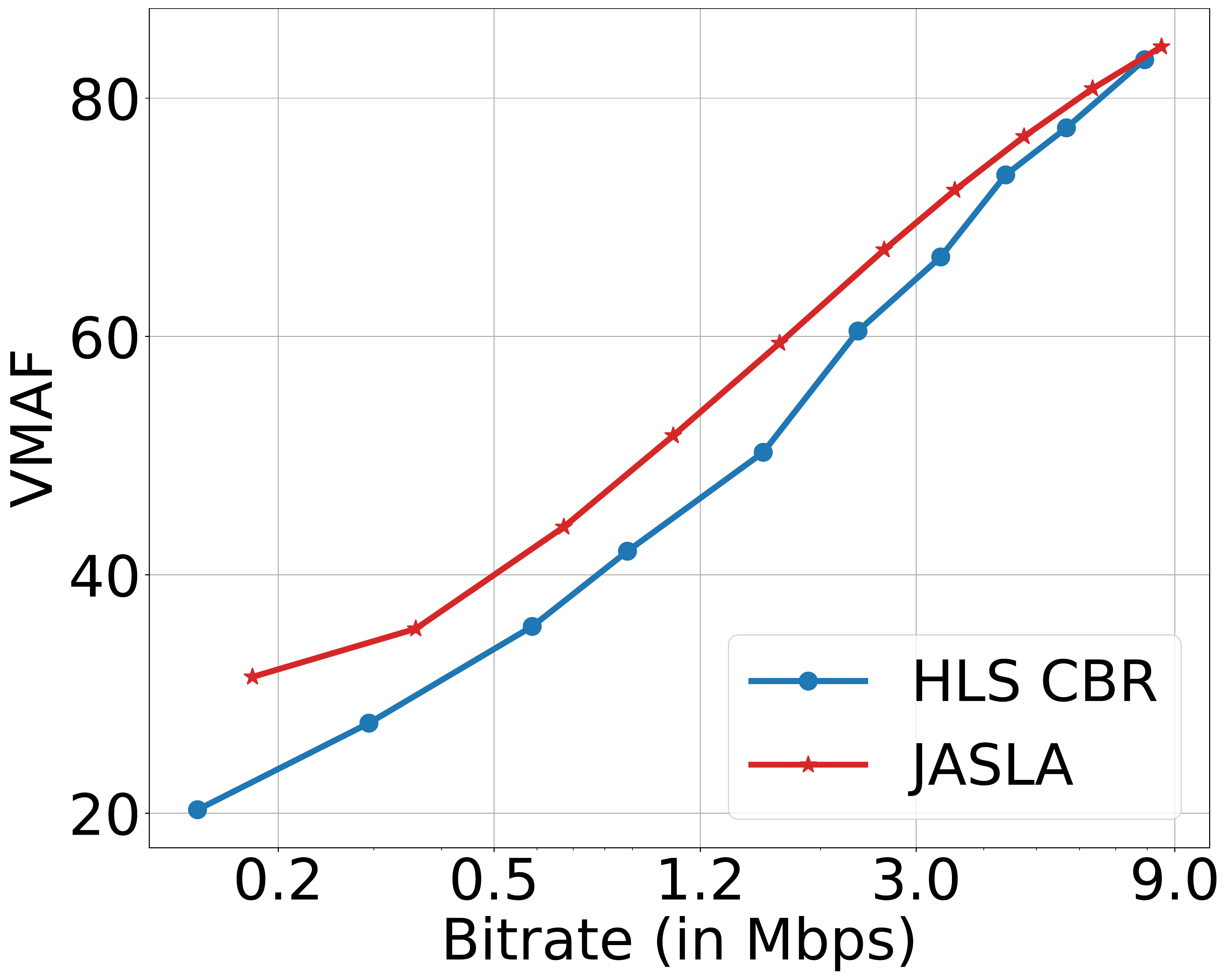}
    \caption{\textit{Runners\_s000}}
    \label{fig:runners_rd1}
\end{subfigure}
\hfill
\begin{subfigure}{0.229\textwidth}
    \centering
   \includegraphics[clip,width=\textwidth]{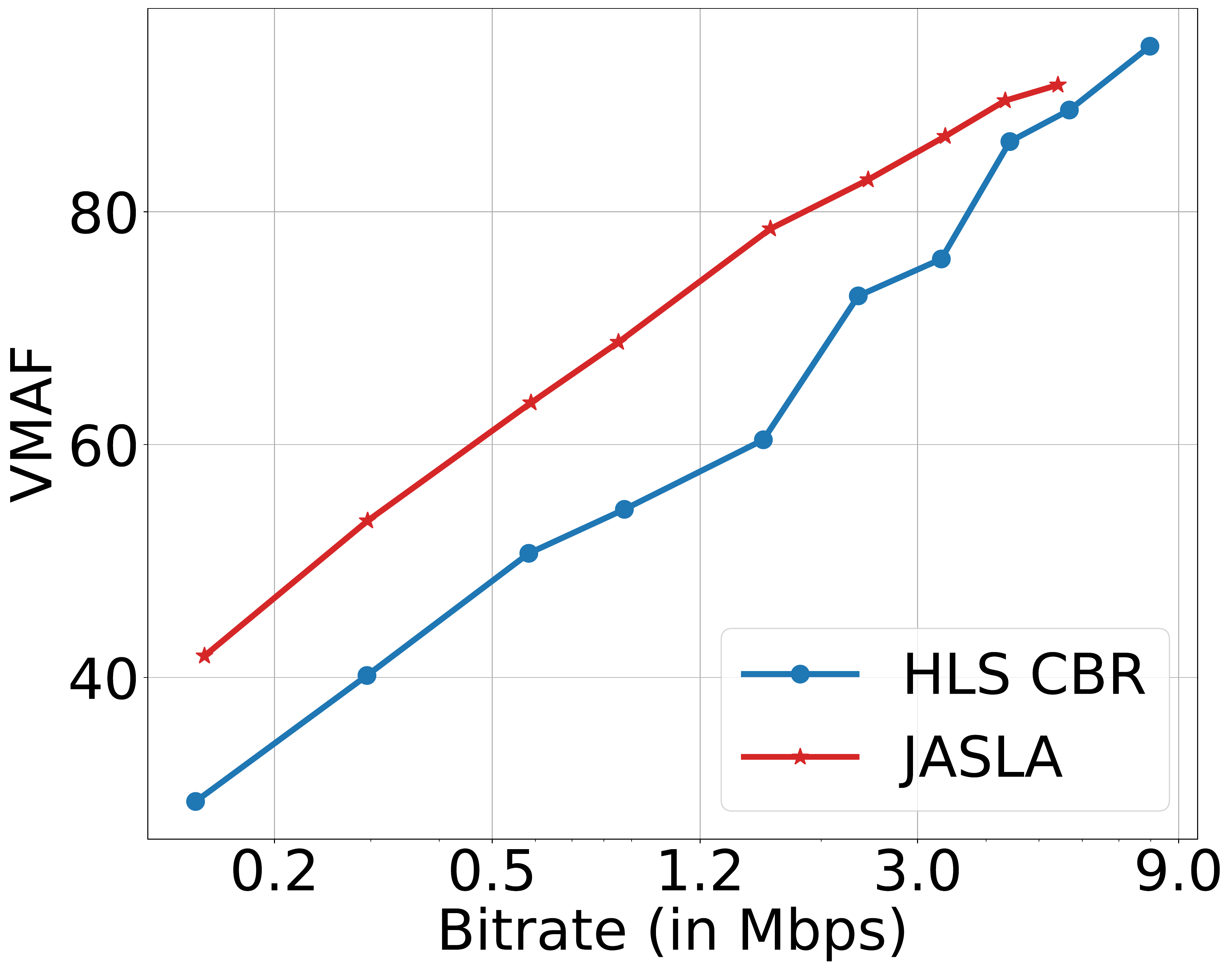}
    \caption{\textit{Wood\_s000}}
    \label{fig:wood_rd1}
\end{subfigure}
\vspace{-0.3em}
\caption{Comparison of RD curves of representative scenes (a) \textit{Bunny\_s000} ($E_{Y}=$22.40, $h$=4.70, $L_{Y}$=129.21), (b) \textit{Bosphorus\_s000} ($E_{Y}$=26.77, $h$=16.08, $L_{Y}$=140.54), (c) \textit{HoneyBee\_s000} ($E_{Y}$=42.93, $h$=7.91, $L_{Y}$=103.00), (d) \textit{RushHour\_s000} ($E_{Y}=$47.75, $h$=19.70, $L_{Y}$=101.66), (e) \textit{Characters\_s000} ($E_{Y}=$45.42, $h$=36.88, $L_{Y}$=134.56), (f) \textit{Eldorado\_s005} ($E_{Y}=$100.37, $h$=9.23, $L_{Y}$=109.06), (g) \textit{Runners\_s000} ($E_{Y}=$105.85, $h$=22.48, $L_{Y}$=126.60), (h) \textit{Wood\_s000} ($E_{Y}$=124.72, $h$=47.03, $L_{Y}$=119.57) using HLS CBR encoding (blue line), \jaslad encoding (red line).}
\vspace{-0.80em}
\label{fig:rd_res_jasla}
\end{figure*}

The following metrics are considered during the evaluation: \textit{(i)} quality in terms of PSNR and VMAF$^{\ref{ref_vmaf}}$, \textit{(ii)} bitrate, and \textit{(iii)} encoding time. Since the content is assumed to be displayed at Full HD (1080p) resolution~\cite{netflix_paper}, the encoded content is scaled to 1080p resolution, and VMAF and PSNR are calculated. Bjøntegaard delta rates~\cite{DCC_BJDelta} $BDR_{P}$ and $BDR_{V}$ refer to the average increase in bitrate of the representations compared with that of the fixed bitrate ladder encoding to maintain the same PSNR and VMAF, respectively. BD-PSNR and BD-VMAF refer to the average increase in PSNR and VMAF, respectively, at the same bitrate compared with the reference bitrate ladder encoding scheme. The relative difference in the storage space required to store all representations ($\Delta S$) is also evaluated as: 
\begin{equation}
    \Delta S = \frac{\sum b_{opt}}{\sum b_{ref}} - 1 
\end{equation}
where $\sum b_{ref}$ and $\sum b_{opt}$ represent the sum of bitrates of all representations in the reference bitrate ladder encoding and \jaslad encoding, respectively.
\vspace{-0.85em}
\subsection{Experimental Results}
\label{sec:exp_results}
\vspace{-0.35em}
The performance of the VMAF, CRF, and JND threshold prediction models is investigated in the first experiment. The average $R^{2}$ score of the VMAF and CRF prediction models are estimated as 0.93 and 0.97, respectively. Hence, a strong positive correlation exists between the predicted and ground truth values. The average MAE of the prediction models is estimated as 3.25 and 1.86, respectively. The MAE of the JND threshold prediction model is observed to be 0.96, which shows that \jaslad works with sufficient prediction accuracy. 

The second experiment analyzes the runtime complexity of \jaslad. \jaslad predicts resolution and CRF at a rate of 300 frames per second, \ie 0.4s per video segment. Compared to~\cite{zhu2022subjective}, the JND prediction runtime in \jaslad is decreased by 97.24\%. 

The third experiment analyzes the bitrate saving and storage reduction results of \jaslad compared to the HLS CBR encoding. Using \jaslad encoding, $BDR_{P}$, $BDR_{V}$, and $\Delta S$ are observed as -34.42\%, -42.67\% and -54.34\%, respectively, compared to the HLS CBR encoding. Moreover, \jaslad encoding yields an average BD-PSNR and BD-VMAF of 2.90 dB and 9.51, respectively. Fig.~\ref{fig:rd_res_jasla} shows the RD curves of eight representative video sequences (scenes) with HLS CBR encoding and \jaslad encoding. The representative scenes exhibit a variety of spatial and temporal complexities (in terms of $E_{Y}$, $h$, and $L_{Y}$). \jaslad yields the highest VMAF at the same target bitrates for all scenes. Moreover, the perceptually lossless representations are eliminated from the bitrate ladder.
\vspace{-1.99em}
\section{Conclusions}
\label{sec:conclusion_future_dir}
\vspace{-0.6em}
This paper proposes a JND-aware per-scene bitrate ladder prediction scheme (\jaslad) for adaptive video-on-demand streaming applications. \jaslad predicts the optimized resolution and corresponding CRF for given target bitrates for every video scene based on content-aware spatial and temporal complexity features. A JND threshold prediction scheme is proposed, eliminating representations that yield distortion lower than one JND from the bitrate ladder. The performance of \jaslad is analyzed using the x265 open-source HEVC encoder against a standard HLS bitrate ladder with the maximum resolution of Full HD (1080p). It is observed that, on average, streaming using \jaslad requires 34.42\% and 42.67\% fewer bits to maintain the same PSNR and VMAF, respectively, compared to the reference HLS bitrate ladder, along with a 54.34\% cumulative decrease in the storage space needed to store representations.

\jaslad shall be extended in the future by preparing a JND prediction model for Ultra HD (2160p) videos, thereby enabling the use of \jaslad in UHD adaptive streaming.
\vspace{-0.6em}
\section{Acknowledgment}
\vspace{-0.6em}
The financial support of the Austrian Federal Ministry for Digital and Economic Affairs, the National Foundation for Research, Technology and Development, and the Christian Doppler Research Association is gratefully acknowledged. Christian Doppler Laboratory ATHENA: \url{https://athena.itec.aau.at/}.

\setlength{\parskip}{0pt}
\setlength{\itemsep}{0pt}
 
\bibliographystyle{IEEEtran}
{\linespread{0.4}\selectfont\bibliography{references.bib}}

\begin{thebibliography}{10}
\providecommand{\url}[1]{#1}
\csname url@samestyle\endcsname
\providecommand{\newblock}{\relax}
\providecommand{\bibinfo}[2]{#2}
\providecommand{\BIBentrySTDinterwordspacing}{\spaceskip=0pt\relax}
\providecommand{\BIBentryALTinterwordstretchfactor}{4}
\providecommand{\BIBentryALTinterwordspacing}{\spaceskip=\fontdimen2\font plus
\BIBentryALTinterwordstretchfactor\fontdimen3\font minus
  \fontdimen4\font\relax}
\providecommand{\BIBforeignlanguage}[2]{{%
\expandafter\ifx\csname l@#1\endcsname\relax
\typeout{** WARNING: IEEEtran.bst: No hyphenation pattern has been}%
\typeout{** loaded for the language `#1'. Using the pattern for}%
\typeout{** the default language instead.}%
\else
\language=\csname l@#1\endcsname
\fi
#2}}
\providecommand{\BIBdecl}{\relax}
\BIBdecl

\bibitem{index2019cisco}
Cisco, ``{Cisco Visual Networking Index: Forecast and Trends, 2017–2022},''
  \emph{White Paper}, February 2019.

\bibitem{Bentaleb2019}
A.~{Bentaleb} \emph{et~al.}, ``{A Survey on Bitrate Adaptation Schemes for
  Streaming Media Over HTTP},'' \emph{IEEE Communications Surveys Tutorials},
  vol.~21, no.~1, pp. 562--585, 2019.

\bibitem{netflix_paper}
J.~{De Cock} \emph{et~al.}, ``Complexity-based consistent-quality encoding in
  the cloud,'' in \emph{IEEE International Conference on Image Processing
  (ICIP)}, 2016, pp. 1484--1488.

\bibitem{VCD_ref}
H.~Amirpour \emph{et~al.}, ``{VCD: Video Complexity Dataset},'' in
  \emph{Proceedings of the 13th ACM Multimedia Systems Conference}, 2022.

\bibitem{jtps_ref}
\BIBentryALTinterwordspacing
V.~V. Menon \emph{et~al.}, ``{JND-aware Two-pass Per-title Encoding Scheme for
  Adaptive Live Streaming},'' 2023. [Online]. Available:
  \url{https://www.techrxiv.org/articles/preprint/JND-aware_Two-pass_Per-title_Encoding_Scheme_for_Adaptive_Live_Streaming/22256704}
\BIBentrySTDinterwordspacing

\bibitem{gnostic}
A.~V. {Katsenou} \emph{et~al.}, ``Content-gnostic bitrate ladder prediction for
  adaptive video streaming,'' in \emph{Picture Coding Symposium (PCS)}, 2019.

\bibitem{res_pred_ref1}
M.~Bhat \emph{et~al.}, ``{Combining Video Quality Metrics To Select
  Perceptually Accurate Resolution In A Wide Quality Range: A Case Study},'' in
  \emph{2021 IEEE International Conference on Image Processing (ICIP)}, 2021,
  pp. 2164--2168.

\bibitem{rep_num_ref}
T.~Huang \emph{et~al.}, ``{Deep Reinforced Bitrate Ladders for Adaptive Video
  Streaming}.''\hskip 1em plus 0.5em minus 0.4em\relax New York, NY, USA:
  Association for Computing Machinery, 2021, p. 66–73.

\bibitem{jnd_ref1}
H.~Wang \emph{et~al.}, ``Prediction of satisfied user ratio for compressed
  video,'' in \emph{2018 IEEE International Conference on Acoustics, Speech and
  Signal Processing (ICASSP)}, 2018, pp. 6747--6751.

\bibitem{mask_effect_ref}
S.~Hu \emph{et~al.}, ``{Compressed image quality metric based on perceptually
  weighted distortion},'' \emph{IEEE Transactions on Image Processing},
  vol.~24, no.~12, pp. 5594--5608, 2015.

\bibitem{jnd_pred2}
Y.~Zhang \emph{et~al.}, ``Deep learning based just noticeable difference and
  perceptual quality prediction models for compressed video,'' \emph{IEEE
  Transactions on Circuits and Systems for Video Technology}, vol.~32, no.~3,
  pp. 1197--1212, 2022.

\bibitem{jnd_ref2}
J.~Zhu \emph{et~al.}, ``{On The Benefit of Parameter-Driven Approaches for the
  Modeling and the Prediction of Satisfied User Ratio for Compressed Video},''
  in \emph{IEEE International Conference on Image Processing (ICIP)}, 2022, pp.
  4213--4217.

\bibitem{zhu2022subjective}
{J. Zhu} \emph{et~al.}, ``{Subjective test methodology optimization and
  prediction framework for Just Noticeable Difference and Satisfied User Ratio
  for compressed HD video},'' in \emph{2022 Picture Coding Symposium}, 2022.

\bibitem{3d_cnn_vqa_ref}
J.~You and J.~Korhonen, ``{Deep Neural Networks for No-Reference Video Quality
  Assessment},'' in \emph{2019 IEEE International Conference on Image
  Processing (ICIP)}, 2019, pp. 2349--2353.

\bibitem{mmsp_paper_ref}
V.~V. Menon \emph{et~al.}, ``{INCEPT: Intra CU Depth Prediction for HEVC},'' in
  \emph{2021 IEEE 23rd International Workshop on Multimedia Signal Processing
  (MMSP)}, 2021, pp. 1--6.

\bibitem{ppte_ref}
{{{V. V. Menon}}} \emph{et~al.}, ``Perceptually-aware per-title encoding for
  adaptive video streaming,'' in \emph{2022 IEEE International Conference on
  Multimedia and Expo (ICME)}, 2022, pp. 1--6.

\bibitem{vqa_mhv_ref}
\BIBentryALTinterwordspacing
{V. V. Menon} \emph{et~al.}, ``{Transcoding Quality Prediction for Adaptive
  Video Streaming},'' 2023. [Online]. Available:
  \url{https://arxiv.org/abs/2304.10234}
\BIBentrySTDinterwordspacing

\bibitem{vca_ref}
{{V. V. Menon}} \emph{et~al.}, ``{Green Video Complexity Analysis for Efficient
  Encoding in Adaptive Video Streaming},'' in \emph{First International ACM
  Green Multimedia Systems Workshop (GMSys '23)}, 2023.

\bibitem{dct_ref}
N.~B. Harikrishnan \emph{et~al.}, ``Comparative evaluation of image compression
  techniques,'' in \emph{2017 International Conference on Algorithms,
  Methodology, Models and Applications in Emerging Technologies (ICAMMAET)},
  2017, pp. 1--4.

\bibitem{ds_paper_ref}
V.~V. Menon, ``{Video Coding Enhancements for HTTP Adaptive Streaming},'' in
  \emph{Proceedings of the 30th ACM International Conference on Multimedia},
  2022, p. 6905–6909.

\bibitem{vqa_icip_ref}
\BIBentryALTinterwordspacing
V.~V. Menon \emph{et~al.}, ``{Video Quality Assessment with Texture Information
  Fusion for Streaming Applications},'' 2023. [Online]. Available:
  \url{https://arxiv.org/abs/2302.14465}
\BIBentrySTDinterwordspacing

\bibitem{bitstream_analyze_ref}
A.~Raake \emph{et~al.}, ``{A bitstream-based, scalable video-quality model for
  HTTP adaptive streaming: ITU-T P.1203.1},'' in \emph{2017 Ninth International
  Conference on Quality of Multimedia Experience (QoMEX)}, 2017, pp. 1--6.

\bibitem{bitstream_analyze_ref1}
R.~R.~R. Rao \emph{et~al.}, ``{Bitstream-Based Model Standard for 4K/UHD: ITU-T
  P.1204.3 — Model Details, Evaluation, Analysis and Open Source
  Implementation},'' in \emph{2020 Twelfth International Conference on Quality
  of Multimedia Experience (QoMEX)}, 2020, pp. 1--6.

\bibitem{GLCM_ref}
R.~M. Haralick \emph{et~al.}, ``Textural features for image classification,''
  \emph{IEEE Transactions on Systems, Man, and Cybernetics}, vol. SMC-3, no.~6,
  pp. 610--621, 1973.

\bibitem{ferri1994comparative}
F.~J. Ferri, P.~Pudil, M.~Hatef, and J.~Kittler, ``Comparative study of
  techniques for large-scale feature selection,'' in \emph{Machine Intelligence
  and Pattern Recognition}.\hskip 1em plus 0.5em minus 0.4em\relax Elsevier,
  1994, vol.~16, pp. 403--413.

\bibitem{x86_simd_ref}
\BIBentryALTinterwordspacing
P.~K. Tiwari \emph{et~al.}, ``{Accelerating x265 with Intel{\textregistered}
  Advanced Vector Extensions 512},'' \emph{White Paper on the Intel Developers
  Page}, 2018. [Online]. Available:
  \url{https://www.intel.com/content/dam/develop/external/us/en/documents/mcw-intel-x265-avx512.pdf}
\BIBentrySTDinterwordspacing

\bibitem{DCC_BJDelta}
G.~{Bjontegaard}, ``{Calculation of average PSNR differences between
  RD-curves},'' \emph{VCEG-M33}, 2001.

\end{thebibliography}
\balance
\end{document}